# A Secure Electronic Prescription System Using Steganography with Encryption Key Implementation

Adebayo Omotosho[*], Omotanwa Adegbola
Department of Computer Science, Bells University of Technology, Ogun State, Nigeria.
[*]Email: bayotosho {at} gmail.com

Olaniyi Olayemi Mikail
Department of Computer Engineering Department of Computer Science and Technology
Federal University of Technology Minna, Nigeria

Justice Emuoyibofarhe
Ladoke Akintola University of Technology
Oyo State, Nigeria.

*Abstract*— **Over the years health care has seen major improvement due to the introduction information and communication technology with electronic medical prescription being one the areas benefiting from it. Within the overall context of protection of health care information, privacy of prescription data needs special treatment. This paper presents an e-prescription system that addresses some challenges pertaining to the prescription privacy protection in the process of drug prescription. The developed system uses spread spectrum image steganography algorithm with Advanced Encryption Standard (AES) key implementation to provide a secure means of delivering medical prescription to the parties involved. The architecture for encoding and decoding was implemented with an electronic health record. The software development tools used were PHP and MySQL database management system for front end and backend data management respectively. The designed system demonstration shows that the synergistic combination of steganography and cryptography technologies in medical prescription is capable of providing a secure transmission to properly provide security for patient's medical prescription.**

## I. INTRODUCTION

The advent of the electronic-age and the internet brought about the use of computer-aided medical support such as, e-health, e-health record and e-prescription. With this information now available electronically, it makes it easier for eavesdroppers, hackers and other malicious attackers to access the confidential information. Connecting personal health information to the internet exposes this information to more hostile attacks compared to the paper-based medical records. Electronic prescribing is simply an electronic way to generate and transmit prescriptions and prescription related information using electronic media between a prescriber and dispensing pharmacy [4]. Electronic prescribing allows a healthcare provider to send accurate and legible prescriptions electronically to a pharmacy. This would minimize medication errors due to the misinterpretation of prescriptions and would eliminate the problem of patients having lost prescriptions.

Patients' health information such as prescription information and health records has always had privacy and confidentiality concerns. The transfer of medical prescriptions from the hospital to a pharmacy is a process that is prone to attack by intruders; the prescriptions may be intercepted, modified or fabricated. The reason why most intruders are so successful is that most of the information they acquire from the system is in a form that they can read and understand. Intruders may reveal the information to others, modify it to misrepresent an individual or use it to launch an attack [1]. Medical prescription and associated medical records which contain sensitive and personal confidential information are transmitted between hospitals and pharmacies on daily basis and are exposed to potential attacks from unauthorized persons. Due to the sensitive nature of the information being transmitted, the possibility of this information being intercepted is a risk which cannot be overemphasized.

While cryptography is all about scrambling of information to an unreadable form, steganography on the other hand is the technique of hiding information in digital media. It is the art and science of hiding data into different carrier files such as text, audio, images, video etc. The main goal of steganography is to communicate securely in a completely undetectable manner and to avoid drawing suspicion to the transmission of a hidden data. In contrast to cryptography, steganography is not to keep others from knowing the hidden information but is to keep others from thinking that the information even exists. To ensure the secure transfer of prescriptions between the prescribing and dispensing systems, this paper presents the development of an electronic prescription system and implements a steganography and a encryption algorithm to secure the prescriptions generated by a doctor or other healthcare provider and ensure only registered pharmacies have the capability to access these prescriptions.

## II. RELATED WORK

Though the advent and development of data transfer through the internet has made it easier to send prescriptions from the hospital to the pharmacy faster and more accurately, it may also be easier for the personal and confidential information to be hacked or stolen in many ways and are vulnerable to attackers who may interrupt, intercept, modify, and fabricate medical prescriptions for various reasons including obtaining drugs for recreational purposes or to feed an addiction or using the prescriptions for medical or financial identity theft and various scandals. Steganography technologies could provide information hiding if properly implemented and would effectively ameliorate the rising demand for prescription information confidentiality and privacy. Some notable works are therefore reported with respect to this:

In [15], the author developed an algorithm, F5, which implements matrix encoding and permutative straddling to improve the efficiency of embedding hidden messages into an image by reducing the number of necessary changes and uniformly spread out the changes over the whole cover image to equalize the embedding rate. The F5 algorithm preserves characteristic properties and offers resistance against visual and statistical attacks. With respect to e-prescription and cryptography, [10] developed an electronic prescription system which uses a public key infrastructure. Confidentiality is ensured by using a symmetric key cryptography, and the introduction of attribute certificates which allow for the allocation of privileges. The electronic prescription is generated and is signed digitally by entering a secret PIN. Then, the prescription is uploaded to a database where it is referenced by its prescription unique identifier where the pharmacy accesses it from. At the pharmacy, the prescriptions is retrieved, the prescriber's signature is verified the prescription is decoded. However, ciphered information may still be easily spotted as been encrypted and an attempt could be make to decipher the content. A steganography algorithm improvement was introduced in [1] in which the authors developed a steganography system which focused on the Least Significant Bit (LSB) technique of hiding messages in an image. The system enhanced the LSB technique to provide a means of secure communication by randomly dispersing the bits of the message in the cover image in contrast to the sequence-mapping technique. The locations of the image pixels in which to embed the secret message are determined by discrete logarithm calculation. Also, a stego key is used in the embedding process making it harder for unauthorized people to extract the original message. The drawback to this is that LSB method of insertion is extremely vulnerable to image manipulation.

Similar to [10], adopting cryptography, [14] described an electronic prescribing system in which a password is required to grant user access to the system. It incorporates public key infrastructure (PKI) technology and digital signatures. Using a PKI, the prescriptions are encrypted before transmission and the receiver must have both a public and a private key to decode the message. Although the system offered high levels of security, privacy and authentication, the limitation associated with this design was the high cost and difficulty in implementing the technology. Likewise, [8] described the use of an electronic prescription system which uses fingerprint recognition as the authentication mechanism. Privacy is ensured by using a public key cryptography algorithm to encrypt the prescriptions and digital signatures are used to bind a public key to a user's identity. The pharmacist, at the receiving end, decrypts and retrieves the prescriptions using the decrypting algorithm and verifies who the prescriptions came from using the digital signature. [7] later developed an electronic prescribing system to increase the efficiency of the prescribing process. The system was created within a web environment which allows the physician to access the prescription system via the internet. The prescriptions would be uploaded by the doctors onto a database where the pharmacy would access it. The limitations the system had were the lack of any security features such as cryptography, steganography, digital certification, firewalls and secure protocols.

Over reliant on cryptography alone seems insufficient; an effort by [2] developed a system which combines steganography with cryptography. They developed an algorithm in which the original message is first converted, letter by letter, to ASCII, and then encrypted using asymmetric key cryptography. After encryption, a steganography algorithm is run to map the bits of the encrypted secret message to pixels of the cover image. The PSNR (Peak Signal to Noise Ratio) and MSE (Mean Square Error) of the stego images produced were examined and proved to have acceptable levels. More on steganography was reported in [11] who developed a steganography system using discrete wavelet transform to increase the hiding capacity and security of the system. The cover image and secret message are normalized and the wavelet coefficient is obtained by applying discrete wavelet transform. The wavelet coefficients of both the cover and message are then fused into a single image to produce a stego image. This method increases the hiding capacity of the system such that the size of the hidden message could be twice the size of the cover image. The experimental results of the algorithm were evaluated based on the PSNR, MSE and entropy between the cover image and the stego image which shows improved values compared to existing algorithms. Furthermore, [6] developed a steganography system to hide data inside an image. They developed an algorithm in which the system uses binary codes inside the pixels of an image. The secret message is converted to binary codes then embedded into the pixels of an image producing a stego image. The stego images were tested using PSNR (Peak signal-to-noise ratio). Based on the PSNR values obtained from tests, it showed that the stego images have quality images without compromising the original image. Also, In [12], a system was developed to secure an electronic voting system using biometrics, cryptography and steganography. The user is authenticated using fingerprint recognition software. Once the vote is casted, it is encrypted using a public key cryptography algorithm and then is embedded in a picture using an LSB insertion algorithm to hide the fact that an encrypted message is being transmitted. The vote is extracted from the image and decrypted using the stegano-cryptography algorithm. Though it uses high security by combining steganography with cryptography, the steganography technique used is weak against statistical attacks. [5] presented the use of biometrics to secure electronic prescription. The prescription system was only secured with physician biometrics and they also proposed a framework whereby a patient can identify himself or herself on the dispensing system via biometrics. However, the work did not consider the security of electronic prescription while in transit.

Spread spectrum technique is similar to another hiding technique that hides data by extending the secret message and placing it over the cover image such as Statistical steganography techniques, transform domain techniques, distortion techniques and least significant bit substitution. The major advantage of spread spectrum is its robustness. Since the encoded information is spread over a wide frequency band, it is difficult to remove it completely without destroying the cover image making it resistant to image manipulation. Also, AES performs three steps on every block (128 bits) of plaintext. Within Step 2, multiple rounds are performed depending on the key size: a 128-bit key performs 9 rounds, a 192-bit key performs 11 rounds, and a 256-bit key, known as AES-256, uses 13 rounds. Within each round, bytes are substituted and rearranged, and then special multiplication is performed based on the new arrangement. AES is designed to be secure well into the future. To date, no attacks have been successful against AES [3].

This paper established the benefits of electronic prescriptions to all the parties involved in the process and the potential to which steganography with AES encryption can be used in securing electronic prescriptions.

### III. METHODOLOGY

#### A. Proposed Electronic Prescription System Architecture

The architecture of the e-prescription system shown in Figure1 is client-server architecture which is a way of sharing resources in which some application programs function as information providers (servers), while other application programs function as information receivers (clients). At least one system is designated as the server which is responsible for performing administrative functions as well as managing and storing the resources of the other systems which are known as the clients. This architecture is preferable in e-prescription because of the large number of users requiring the same service and information, hence, a central server can service the many clients involved.

The electronic prescribing system has three phases which includes; **the Registration phase**, the **Prescribing phase** and the **Dispensing phase**. This architecture provides the application high flexibility and improved efficiency. **The Registration phase** involves the registration of all the entities that will enable the outcome of the prescription process such as the prescribers, the dispensers and administrators. The administrators are responsible for updating the list of current medications available, providing a list of registered pharmacies, creating and granting authorization to different users; the doctors are authorized to prescribe medications and upload them to the server, while the pharmacists are authorized to view the prescriptions and dispense the medications. **The Prescribing phase** involves the process of actually prescribing medication and uploading the prescriptions to a server. The prescriber is authenticated using the registered username and password before granted access to use the system. To write a prescription, the prescriber first identifies the patient from the electronic health records (EHR) which allows the prescriber to review the patient's data, medication history, checks for allergies and potential drug interactions before generating the prescription. After prescribing the medications, a steganography algorithm is implemented, which is specifically the spread spectrum algorithm, to embed the AES encrypted prescription inside an image before it is sent to the prescription server. **The Dispensing phase** is where the medications are dispensed to the patients. At the pharmacy, the pharmacist logs on to the system with a designated username and password and is granted access to the dispensing interface. The crypto stego image is retrieved from the prescription database and passed through the decrypting steganography algorithm which extracts the prescription form the image if the decryption key is available. Each pharmacy system is capable of accessing the prescription database so the patients can collect the prescribed medication at any pharmacy of choice. After the medication has been dispensed, the pharmacist would then upload information about the dispensed medication back to the database server confirming that the prescription has been filled.

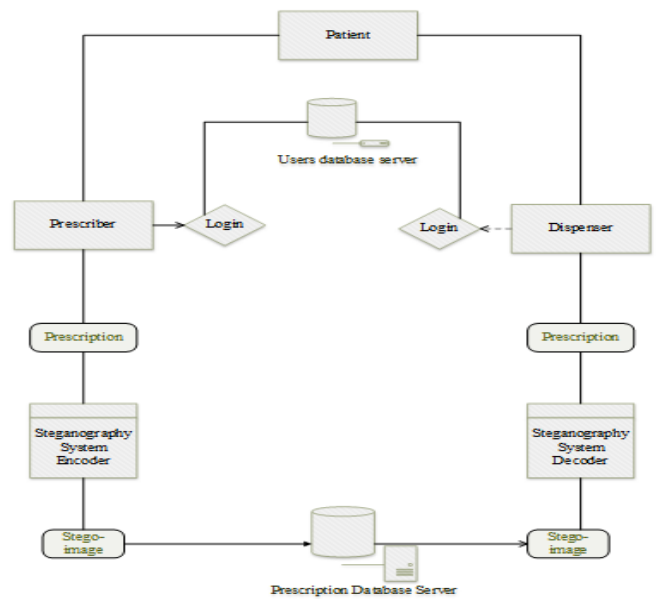

Figure 1: Proposed Electronic Prescription System Architecture

#### B. Spread Spectrum Steganography Technique

The fundamental concept of spread spectrum steganography technique is the embedding of the hidden information within noise, which is then added to the digital image. The system hides a significant number of image bits inside digital images; if the noise is kept at low levels, it is not perceptible to the human eyes or susceptible to detection by computer analysis. The advantages of spread spectrum are its robustness and difficulty in detecting and removing a signal. Since the encoded information is spread over a wide frequency band, it is difficult to remove it completely without destroying the cover image making it resistant to image manipulation. Furthermore, the spread spectrum steganography system is a blind steganography scheme, in the sense that the original image is not needed to extract the hidden information; only a secret key is required. This paper implements the algorithm presented in [13].

### C. Embedding Process

Figure 2 shows spectrum steganography message embedder. The process of embedding hidden messages inside an image consists of the following steps

1. Create encoded message by adding redundancy via error-correcting code.
2. Add padding to make the encoded message the same size as the image.
3. Interleave the encoded message.
4. Generate a pseudorandom noise sequence, n.
5. Use encoded message, m, to modulate the sequence, generating noise, s.
6. Combine the noise with the original image, f.

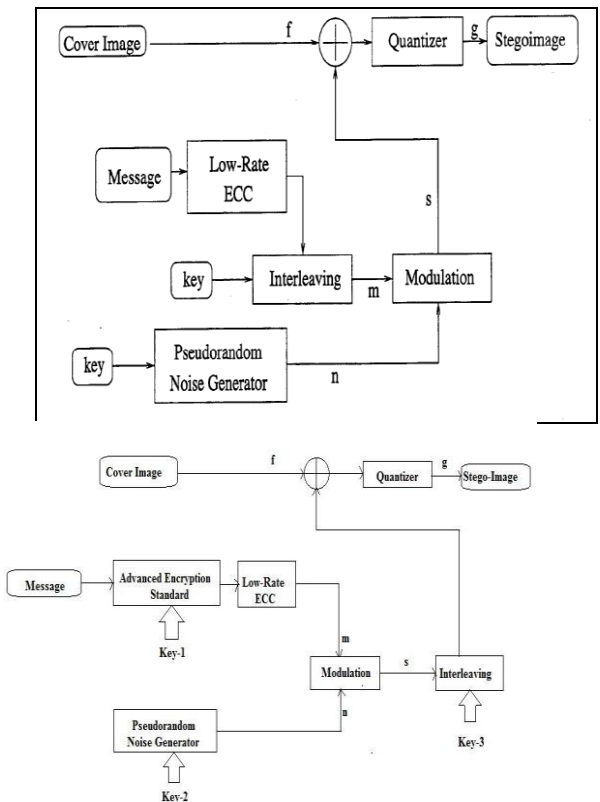

### D. Decoding Process

Figure 3 shows spread spectrum steganography message extractor. In the decoding process, the original image is not required to retrieve the hidden message; a filter is used to extract the noise from the stego-image to produce an approximation of the original image. The process of retrieving the hidden message is as follows:

1. Filter the stego-image, g, to get an approximation of the original image, f^.
2. Subtract the approximation of the original image from the stego-image to get an estimate of the noise, s, added by the embedder.
3. Generate the same pseudorandom noise sequence, n.
4. Demodulate by comparing the extracted noise with the regenerated noise.
5. De-interleave the estimate of the encoded message, m, and remove the padding.
6. Use error-correcting decoder to repair the message as needed.

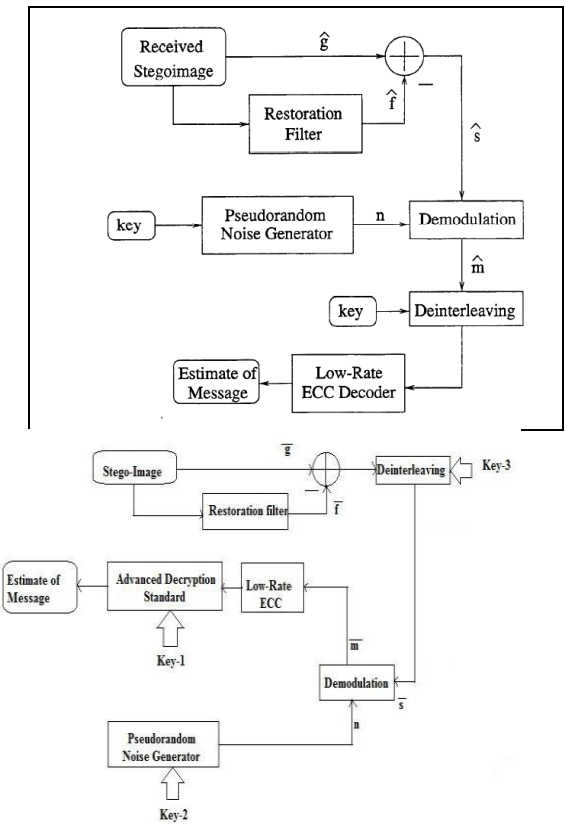

Figure 2: Spread Spectrum Steganography Message Embedder without Cryptography [9] and Spread Spectrum Steganography Message Embedder with Cryptography [13]

Figure 3: Spread Spectrum Steganography Message Extractor without cryptography [9] and Spread Spectrum Steganography Message Extractor with cryptography [13].

## IV. RESULTS AND DISCUSSION

Before you begin to format your paper, first write and save the content as a separate text file. Keep your text and graphic files separate until after the text has been formatted and styled. Do not use hard tabs, and limit use of hard returns to only one return at the end of a paragraph. Do not add any kind of pagination anywhere in the paper. Do not number text heads- the template will do that for you.

*Operation of the electronic prescription system*

The secure electronic prescription system has 3 categories of users; The Prescribers, The Dispensers, and The Administrators. The Prescribers include physicians and other registered healthcare providers who prescribe medications for patients. The Dispensers include registered pharmacists who dispense medications prescribed by the prescribers. The Administrators are in charge of managing the operations of the electronic prescription system which include registration of users, managing the list of registered pharmacies and of accredited drugs. Some of the interfaces involved in its operations are highlighted below:

A. *Login Page*

Figure 4 represents the login page which is the first interface of the electronic prescription system, and it gives the registered users (Prescribers and Dispensers) the means to access the system using their registered username and password. The system is designed to check the user name to determine if the user is a prescriber (physician) or dispenser (pharmacist).

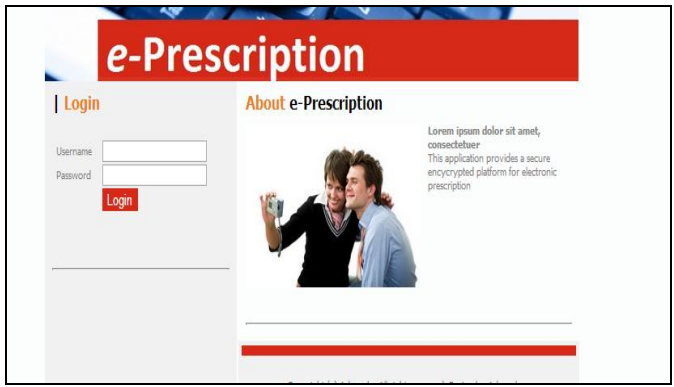

Figure 4: Login page for the Electronic prescription system

B. *Prescribing Page*

The prescribing page enables the prescriber to generate prescriptions and upload it to the database where the pharmacist can access it from. Once a prescription has been generated, it provides the physician with an interface to both advice the patient as well prescribe drugs to help the patient. Figure 5 provides a pictorial view of this.

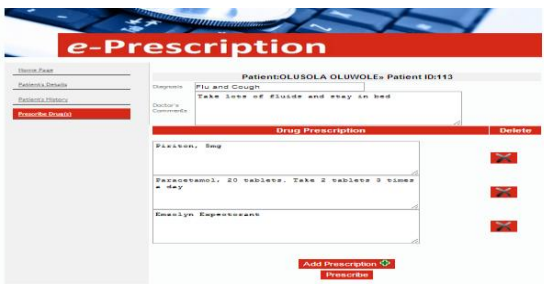

Figure 5: Prescribing Page

C. *Confirmation page*

After prescribing, the confirmation page is shown clarifying that the prescriptions have successfully been saved on the database. The prescriber then clicks on the link "Print" to generate the Prescription Identification Code. Figure 6 represent the page.

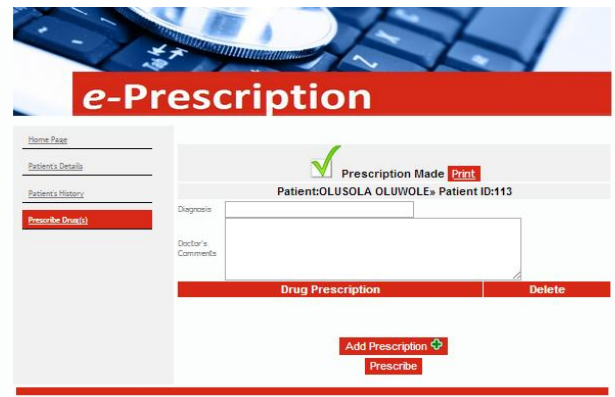

Figure 6: Confirmation Page

D. *Prescription Information Code*

Figure 7 represent the Prescription Identification Code this is generated when a prescriber prints his/her medication. This code is required of the pharmacy during the dispensing process to decode the prescriptions from the image. This code generated code is actually the Advanced Encryption Standard (AES) key for the encrypted e-prescription.

| Doctor | ade |
|---|---|
| Patient | OLUSOLA OLUWOLE |
| Prescription Code | 1131373638606 |
| Date | 2013-7-12 |

Print

Figure 7: Prescription Information Code

E. *Dispensing Page*

The dispensing page as represented by Figure 8 shows the dispenser the medication history of a patient. It shows the prescriptions that have been collected and the prescriptions that are yet to be collected. The pharmacist can dispense drugs to the patients by clicking the "Dispense" link. After dispensing medication, the drug status is updated to reflect that the drugs have been dispensed.

Figure 9 displays the stego image this shows when the dispenser selects a medication to dispense. It requires the dispenser to enter the Prescription Identification Code that was generated during the prescribing process. The prescription is encoded using spread spectrum algorithm and AES encryption.

Figure 10 represents the interface displayed after the dispenser enters the Prescription Identification Code displaying the prescriptions. The dispenser is required to click on the "Dispense" link after dispensing the medications to send a feedback to the server stating that the prescription have been dispensed, this is done in other to notify the physician

Figure 11 represents dispenser's confirmation page. It confirms to the dispenser that the details of the prescriptions have successfully been updated on the database. This is to prevent cases of double dispensing of medications.

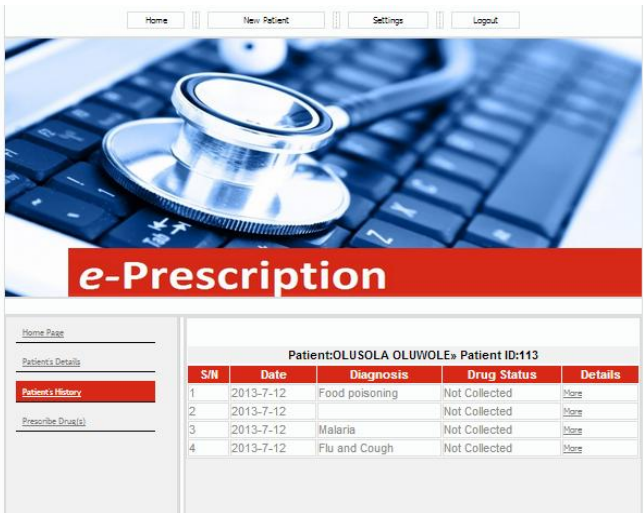

Figure 8: Patient's History

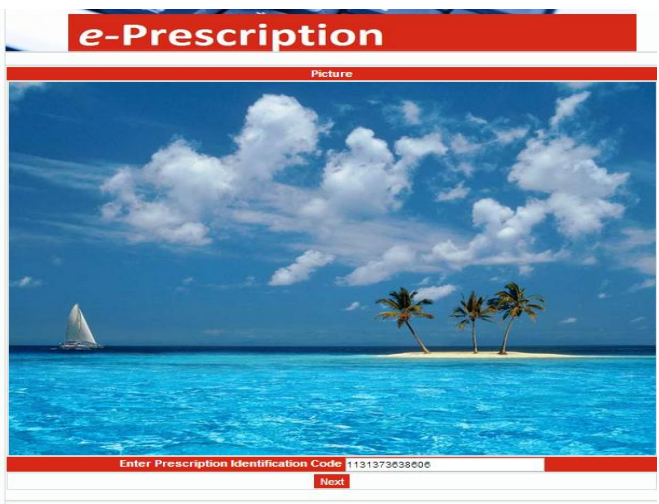

Figure 9: Image with Prescription Encoded Within Using Spread Spectrum Algorithm

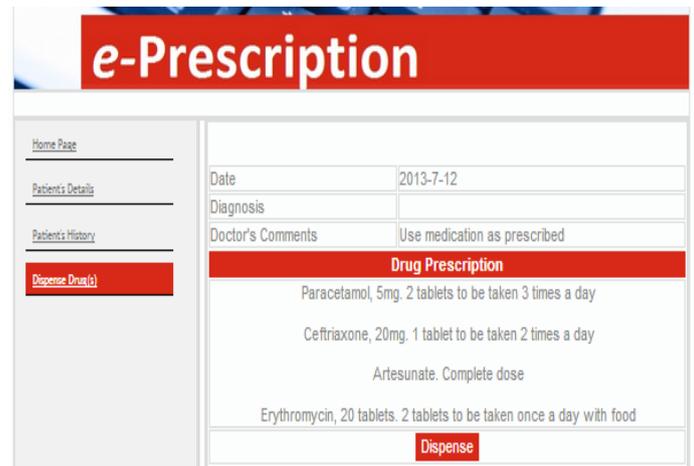

Figure 10: Prescription page

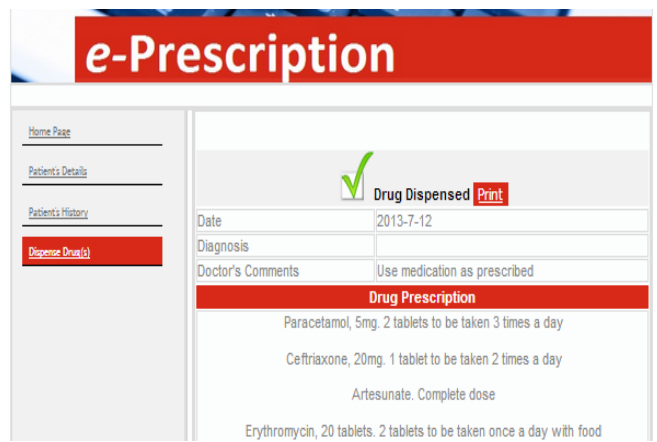

Figure 11: Confirmation page

## V. CONCLUSION

In this paper an approach to developing a secure electronic prescription system has been presented, this makes use of steganography and AES encryption key which is a highly secured method for prescription communication between the prescriber and the pharmacists. The strength of the system lies in the implemented steganography algorithm which ensures effective data hidden from un-authorized access. This guarantees that the issue of security in the e-prescription system and thus makes the clients put their confidence in the developed system. The use of steganography and cryptography creates a strong form of security in data communication. The secured system will ensure a better, easier and more efficient means of prescribing and transmitting medical prescription. Future work should consider using audio and video spread spectrum steganography and performance evaluation of the developed system using appropriate metrics.